\documentclass{jaa} 
\usepackage{natbib}
\usepackage{color}
\usepackage{graphicx}
\usepackage{ulem}
\usepackage{multirow}
\usepackage{array}

\begin{document}

\title{On the simultaneity of Forbush decreases: the simultaneous effects of interplanetary parameters and geomagnetic activity indices}


\author{I. O. Eya\textsuperscript{1,2,3,4*}, E. U. Iyida\textsuperscript{2,4}, O. Okike\textsuperscript{2,4,6}, R. E. Ugwoke\textsuperscript{2,4,5}, F. M. Menteso\textsuperscript{2,4}, 
C. J. Ugwu\textsuperscript{3,4,7}, 
P. Simpemba\textsuperscript{1,}, J. Simfukwe\textsuperscript{1}, D. Silungwe\textsuperscript{1}, S. P. Phiri\textsuperscript{1}, G. F Abbey\textsuperscript{1},
 J. A. Alhassan\textsuperscript{2,4} \and A. E. Chukwude\textsuperscript{2,4}}
\affilOne{\textsuperscript{1}Department of Physics, The Copperbelt University, Jambo Drive, Riverside, Kitwe, 10101, Zambia\\}
\affilTwo{\textsuperscript{2}Astronomy and Astrophysics Research lab, University of Nigeria, Nsukka, Nigeria.\\}
\affilThree{\textsuperscript{3}Physics/Electronics Technique -- Department of Science Laboratory Technology, University of Nigeria, Nsukka, Nigeria.\\}
\affilFour{\textsuperscript{4}Department of Physics and Astronomy, University of Nigeria, Nsukka, Nigeria.\\}
\affilFive{\textsuperscript{5}Department of Physics, Federal University of Technology Owerri, Imo State Nigeria.\\}
\affilSix{\textsuperscript{6} Department of Industrial Physics, Ebonyi State University, Abakaliki, Nigeria.\\}
\affilSeven{\textsuperscript{7}Department of Mathematical Sciences, University of South Africa, Florida Park, 1709, Roodepoort, South Africa\\}


\twocolumn[{

\maketitle

\corres{innocent.eya@unn.edu.ng}


\begin{abstract}
Forbush decreases (Fd) are transient, short-term reductions in the intensity of galactic cosmic rays that reach the Earth's surface.  When this reduction is observed at multiple locations at the same time, it is referred to as simultaneous Forbush decreases (SFd). Identifying Fd events in daily averaged Cosmic ray (CR) raw data is always tedious, but the task has gone minimal through an algorithm (automated Fd detection). We deployed an automated Fd location algorithm on daily-averaged CR data from five neutron monitor stations covering the period 1998 to 2006. We identified 80 days with the most simultaneous events. While there exists extensive research on the subject using a case study approach, the current study is statistical. Whereas most of the previous investigations employed a small sample of Fds manually selected from a single CR station, large samples of Fds, selected after disentangling the Sun's influence on CR data from multiple neutron monitors (MNs) are used. The connection between the Fds and many solar-terrestrial variables is tested. The beautiful and consistent results obtained between the space weather variables and Fds at the five NM stations call the attention of space weather researchers to the need for rigorous, detailed, and accurate cataloging of Fds. Solar cycle oscillation significantly impacts the amplitude and timing of Fds. Its influence should be removed before Fd selection. 
\end{abstract}
}]



\doinum{12.3456/s78910-011-012-3}
\volnum{000}
\year{0000}
\pgrange{1--}
\setcounter{page}{1}
\lp{1}

\section{Introduction}
\label{sect:intro}
High-energy charged particles with energies 100 MeV to  $10^{20}$ eV known as Cosmic rays (CRs) \citep{Yu2015}  originating from outer space do bombard the Earth's surface naturally. 
The interaction of these particles with transient events from the Sun provides information about incoming atmospheric disturbances before they reach the Earth. 
Studying the variation of CR is crucial in achieving a better understanding of the Sun-Earth interaction \citep[e.g:][]{Lockwood1971,Stozhkov2003,Usoskin2004,Belov2006}.
Researchers have found that changes in CR intensity can be linked to various solar phenomena such as coronal mass ejections (CMEs), solar flares, and geomagnetic storms \citep[e.g:][]{Cane2000,Stozhkov2003,Subramanian2009}.
 Additionally, it has been suggested that CR intensity variation is connected to changes in cloudiness \citep{Lockwood1991,Pudovkin1995,Pierce2009}, atmospheric electricity \citep{Marcz1997,Antonova2023},
 temperature, thunderstorms, lightning activity \citep{Gurevich1999,Dorman2005} and depletion of the ozone layer  \citep{Palle2000,Fedulina2001}. 
Studies of CR intensity variation are therefore essential to comprehensively understand the impact of solar activity on our environment.
CMEs and Solar Wind are some of the events that interact with CRs permeating the Earth's surface \citep{Subramanian2009}. 
One way of quantifying the interaction is to measure the changes in CR intensity reaching the Earth. 
Neutron Monitors (NMs) record the changes in CR intensity over time. 
Changes in CR intensity can be categorized as periodic or non-periodic, and both are believed to be influenced by solar activity  \citep{Lockwood1991,Firoz2010}. 
Periodic changes include the CR diurnal anisotropy \citep{Okike2021a,Okike2021b,Alhassan2022a}, as well as 27-day and 11-year long-term modulations \citep{Oh2008}. 
The CR diurnal anisotropy is a short-term variation caused by the 24-hour rotation of the Earth about its axis \citep{Lockwood1971}. 
Non-periodic changes manifest in two phenomena: the Forbush decrease (Fd) and the Ground Level Enhancement (GLE) Fds refer to sudden decreases in CR intensity \citep{Forbush1937,Belov2009}, while GLEs result in an increase in CR intensity
 \citep{McCracken2012,Gopalswamy2012,Hubert2019,Ugwoke2023}. 
Fd events usually comprise four phases, namely, the onset, main phase, point of maximal depression, and recovery phase \citep{Oh2008, Bhaskar2016,Alhassan2022b}. 
The time scale of the duration of the Fd profile can last from several hours to days 
\citep{Alhassan2022b}.

The identification of Fds has predominantly been a manual process since the discovery of the event in 1937 \citep[eg][and references therein]{Forbush1937,Forbush1938,Oh2008,Kristjansson2008,Oh2009,Lee2013,lee2015}. This technique involves downloading CR data of preferred resolution, visually inspecting the plotted data for points of maximum depression, recording the onset and end time dates of the main phase, and calculating individual Fd amplitudes.
However, this method has several limitations \citep[see][for details]{Okike2020a,Okike2020b}, such as its tediousness, lack of accounting for spontaneous signals in the CR data (such as diurnal CR anisotropies), bias and subjectivity arising from trial and error, and inability to analyze many Fds concurrently.
In recent times, an automated approach to Fd identification has emerged as a viable alternative \citep[eg][]{ Ramirez2013,Okike2019,Okike2020a,Light2020a}. The automated method is based on a Fourier transform technique and R algorithm to derive the Fd amplitude and time of minimum depression from raw CR data \citep[see][for details]{Okike2020a,Okike2020b,Alhassan2022a,Alhassan2022b}. This technique filters out spontaneous signals in the CR data, including diurnal CR anisotropies, and both low- and high-frequency superposed signals. The R code then identifies the points of minimum depression and computes their magnitude (Fd) and event time.
This automated technique is more efficient and addresses the limitations of the manual approach. By leveraging this automated approach, researchers can analyze numerous Fds concurrently at many NM stations with ease.

Fds can be classified based on their sizes \citep{Belov2001,Okike2011,Okike2020b}.
It is interesting to note that the size of an Fd is a relative measure with respect to the initial CR count rate. 
So the sign Fd bears is as the researcher chooses.
Events with an absolute size of 3\% or more are often considered to be strong or large Fds \citep{VanAllen1993,Cane1993,Cane1996,Oh2008,Harrison2010,Laken2012}, while those with a size of less than 3\% are considered to be weak or small Fds \citep{Cane1993,Pudovkin1995,Oh2008,Okike2021a}. 
As such, in this analysis events whose absolute value  $ \geq 3 \% $ are considered as large Fds.
  
Solar activity can interact with interplanetary activity, influencing the motion and intensity of CR flux as they travel towards Earth \citep{Badruddin2006,Lingri2016}. 
Such activities that are of interest to us are Solar wind speed (SWS) and interplanetary magnetic field (IMF).  
Solar wind is a stream of charged particles, primarily electrons and protons that are continuously owing from the Sun's corona into the interplanetary space, which on reaching the Earth, leads to geomagnetic activity. 
The speed can vary from about 250 to 700 km/s depending on the activity level of the Sun \citep{S2004} . 
When the SWS is high, it can compress the Earth's magnetic field creating a region called the "bow shock," which can deflect some of the CRs away from the Earth \citep{Badruddin2006}. 
As a result, the intensity of CRs reaching the Earth's surface is reduced. 
Moreover, the solar wind carries a magnetic field permeating the interplanetary space called the interplanetary magnetic field (IMF)\citep{Gosling1997,Raghav2014}. 
When the IMF is oriented in a direction that is parallel to the Earth's magnetic field lines (i.e., "northward" orientation), it can provide additional shielding from CRs, reducing the number of CRs that penetrate the Earth's atmosphere. 
Another phenomenon that influences the IMF to usurp the intensity of the CR is Interplanetary Coronal Mass Ejections (ICMEs)\citep{Cane2000,Badruddin2002,Subramanian2009,Raghav2014} or Interplanetary Shocks (IP shocks) near Earth. 
If the ICMEs are directed towards the Earth, it can interact with the Interplanetary Magnetic Field (IMF) and cause a compression of the magnetic field. 
This compression can block some of the CRs from entering the Earth's atmosphere, resulting in a transitory decrease in their intensity -- Fd \citep{Badruddin2002,Candia2004,Subramanian2009}. 
Meanwhile, the Earth's magnetosphere acts as a shield against CR particles, particularly those with lower energy. 
Any time there is solar activity, there is geomagnetic response of the magnetosphere inform of disturbances \citep{Rangarajan2000}.
The geomagnetic disturbances are weighed by geomagnetic indices \citep{Firoz2010} such as the planetary K-index (kp), planetary A-index (ap), and disturbance storm time index (Dst).
They provide information about the state of the Earth's magnetosphere. 
In this analysis, we shall use these indices to access the state of the Earth's magnetosphere at various locations of the NMs at the time of the Fds. 
The Kp is a measure that indicates the level of global geomagnetic activity (i.e. disturbance of the Earth's magnetic field) caused by the interaction of solar wind with the Earth's magnetosphere \citep{Bartels1939,Matzka2021}. 
It represents the deviation of the Earth's magnetic field from its normal behaviour on a quiet day.
The Ap is a metric that assesses global geomagnetic activity over a 24-hour period. 
The Ap index is obtained through the computation of the Kp index \citep{Matzka2021}, which is predicated on evaluations of magnetic field variations in diverse magnetic observatories worldwide. The Ap index is derived from the daily average of the eight 3-hour Kp indices for a particular day. 
It provides a summary of the level of geomagnetic activity for that day. 
The Dst index is a measure of the intensity of geomagnetic storms \citep{Borovsky2017} unlike the Kp, which provides a global measure of geomagnetic activity. The Dst index specifically represents the strength of the equatorial ring current. 
The equatorial ring current is a band of charged particles, primarily electrons and ions, which encircle the Earth at the geomagnetic equator. 
During geomagnetic storms, an inflow of energetic particles from the solar wind enhances the ring current \citep{Cramer2013}, causing a temporary weakening of the Earth's magnetic field. 
A larger negative value of Dst denotes a stronger geomagnetic storm, whereas positive values represent quiet geomagnetic conditions.

\section{Simultaneity of Forbush decrease}
Fd simultaneity refers to a simultaneous reduction in CR intensity at multiple locations on or near Earth's surface. 
This means that when an Fd event is observed at one location, it is also observed at other locations worldwide. 
The global nature of Fds is crucial because it implies that the reduction in CR intensity is not just a local phenomenon. 
Observing an Fd at multiple locations simultaneously can provide vital information about the event's spatial and temporal characteristics, including its onset time, duration, and magnitude. 
This is important in understanding the global effects of Fds on Earth's atmosphere and space weather.
In analyzing Fds recorded by high latitude NMs (Oulu, Magadan, and Inuvik) \cite{Oh2008} and \cite{Oh2009}  classified Fds into two categories: simultaneous and non-simultaneous. 
They maintain that simultaneous Fds (SFds) occur when the main phase of the CR intensity profile overlaps in universal time (UT), while non-simultaneous Fds occur when the main phase overlaps in local time (LT) \citep{Oh2009}. 
Meanwhile, when a CME interacts with the solar wind, it can create a magnetic cloud \citep{Burlaga1981}. 
That is a region within a CME where the magnetic field lines are highly organized in the form of magnetic flux ropes that have a distinct orientation. 
In that, \cite{Oh2008} noted that the intensity and direction of magnetic clouds determine whether Fds are to be simultaneous or not.  
If IP shocks and intense magnetic clouds stream symmetrically and directly to Earth, it results in SFds. 
 In this case, Earth is the center of the magnetic impediment, and all NMs will record a drop in CR intensity at the same UT (resulting in SFd). 
 However, if Earth is not at the center of the magnetic impediment, only NMs within the stream path may record drops in CR intensity, leading to non-simultaneous Fd.
\cite{Lee2013} using data from middle-latitude stations (Irkutsk, Climax, and Jungfraujoch) during the solar maximum (1998 -- 2002) statistically classified Fd into simultaneous and non-simultaneous.
  In that, it is shown that Fd events at middle latitudes also showed differences in the onset times of the main phases, similar to events observed at high latitudes.

\begin{figure}[h]
	\centering
	\includegraphics[width=\columnwidth]{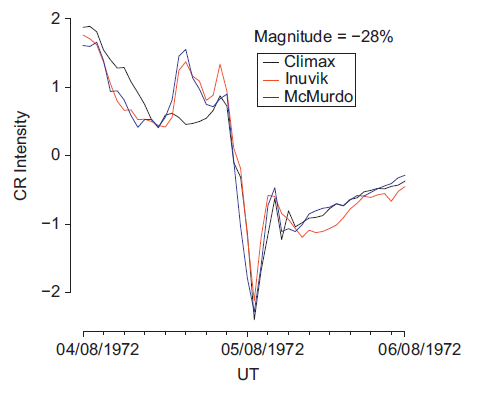}
	\caption{Illustration of simultaneous Fd event} [adopted from  \cite{Okike2011} --- Figure (1)].
	\label{fig:Okike}
	 \end{figure}

 \cite{Okike2011} use the complete profile of Fd events such as the onset, main phase, Fd minimum point, and the recovery phase in identifying simultaneous events [see Figure (\ref{fig:Okike})] and further employed the method of principal component analysis (PCA) to differentiate between globally simultaneous and non-simultaneous Fds. In their report, SFd events observed at all the PC1 signals registered a high correlation (r $ \sim $ 1) with data for each of the stations. In contrast, the PC1 signal registered a very weak correlation $(r < <1)$  with data for each of the stations for the non-simultaneous Fds. Based on the observation of Figure (\ref{fig:Okike}), it can be seen that the maximum intensity of Fds decreases simultaneously at the same universal time for all NMs, regardless of their location. Additionally, there are significant similarities in the structure of both the main and recovery phases of the event at all three stations. These structural similarities suggest that Fd events have a common origin. 

The foregoing suggests that Fd event simultaneity may be defined with regard to some parts of Forbush events. This implies that considering the structure of the Fds is necessary for identifying SFds. Figure \ref{fig:Okike} clearly shows that the time of Fd minimum is unique and shows greater similarity than other parts of the Fd at the three stations. \cite{Okike2020a} defined Fd event simultaneity with reference to the timing of the Fd minimum. An Fd event is taken as globally simultaneous if the event time of minimum happened at the same time irrespective of the location of the NM. In this analysis, we followed the method of \cite{Okike2020a} by timing Fd event using the Fd minimum in identifying the SFds. Data from five CR stations (Calgary (CALG), Oulu(OULU), Apatity(APTY), Novosibirsk(NVBK), and South Pole(SOPO)) are used in the current work. The station locations are indicated in Figure \ref{Fig:location}.

\section{Data and Method}
\subsection{Source of data}
\begin{figure}
\centering
\includegraphics[scale=0.6]{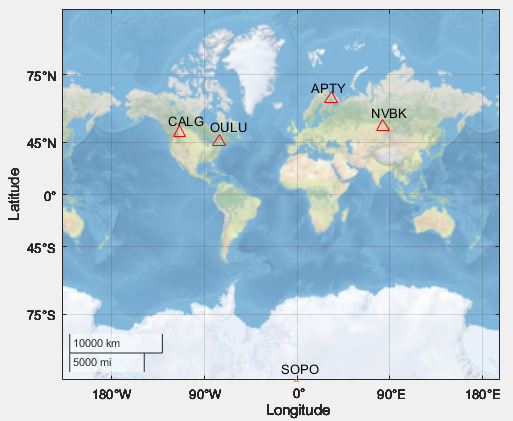}
\caption{Location of the five neutron monitors in a global map}
\label{Fig:location}
\end{figure}

The daily averaged CR  raw data corrected for atmospheric pressure, for the determination of the Fds were downloaded from   
http://cr0.izmiran.ru/common. 
An advantage of using daily averaged CR data is that the effects of CR diurnal anisotropy are reduced \citep{Dumbovic2011,Belov2018}.
The data is between 1998 to 2006 and for the neutron monitor stations presented in Figure \ref{Fig:location}. The parameters of the stations are presented in Table ~(\ref{tb:NM_station}).

The IZMIRAN research team collects data from individual CR stations and stores them in the same format in a common text -- American Standard Code for Information Interchange (ASCII,). 
The data format for all the station is of the form -- yyyy.mm.dd HH:MM, followed by the CR count. 
The corresponding simultaneous daily SWS, IMF, kp, Dst, and Ap data used in the analysis are from https://omniweb.gsfc.nasa.gov/html/ow data.html.

\begin{table*}
\centering
\caption{Parameters of the Neutron monitor stations}
\label{tb:NM_station}
\begin{tabular}{llcccc}
\hline
Full name & Abbreviation &Latitude ($ ^{0} $) & Longitude ($ ^{0} $) & Altitude & Rigidity (GV) \\
\hline
Oulu & OULU (ou) &  45.40 & -75.70  &   0 &  0.77\\
South Pole &SOPO (so) & -90.00 &  0.00   &  2820 &  0.09 \\
Apatity & APTY (ap) &  67.50 &  33.30  &   117 &   0.57\\ 
Calgary & CALG (ca) &  51.10 & -114.10 &  1128 & 1.08 \\
Novosibirsk & NVBK (nv) &  54.80 &  83.00  &  163 & 2.69\\
\hline
\end{tabular}
\end{table*}
 
\subsection{Method of Analysis}\label{Analysis}

\begin{figure}
	\includegraphics[width=\columnwidth]{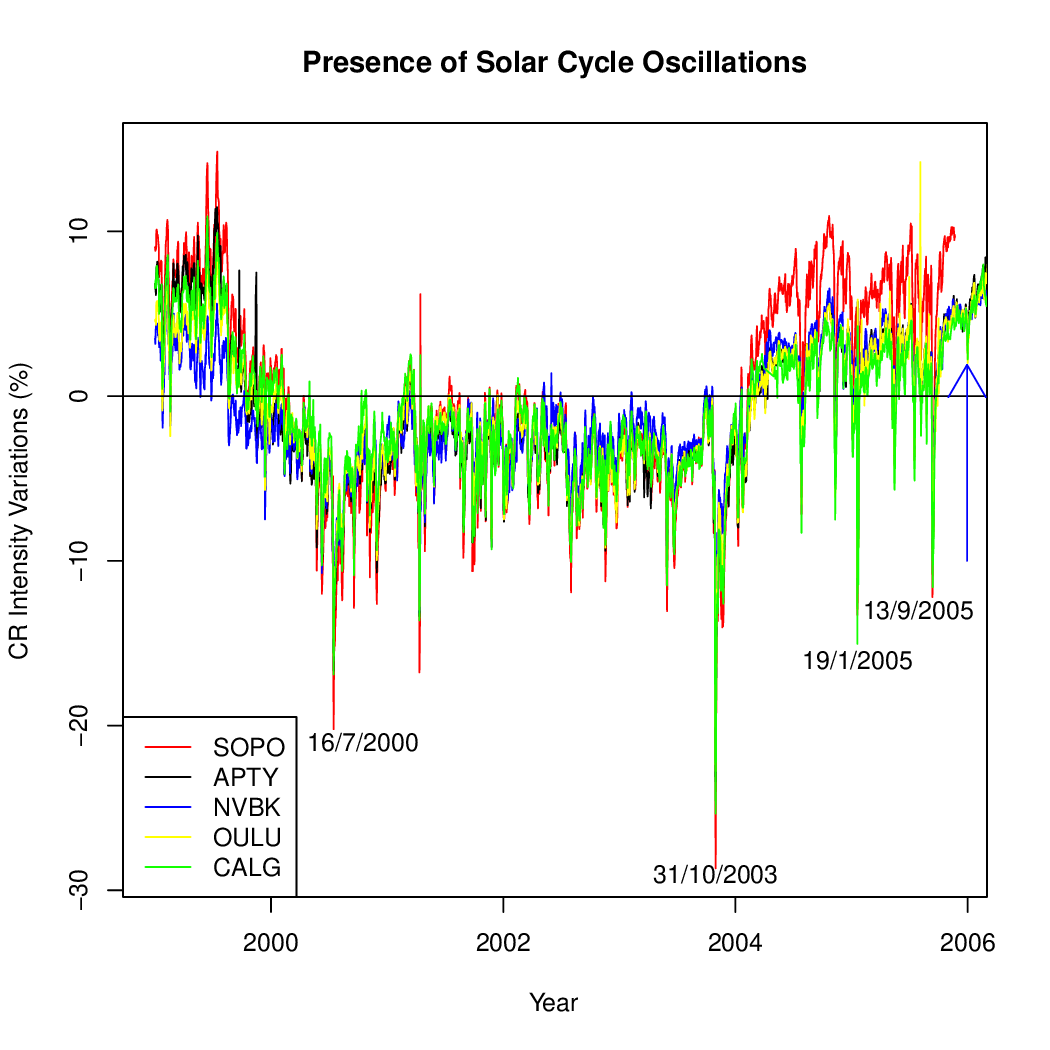}
    \caption{CR data variations showing the presence of solar cycle oscillation at the five CR stations.}
    \label{Figureraw}
\end{figure}

Owing to the absence of a comprehensive Fd list from any of the existing NMs, new catalogs of Fds continue to appear in the literature. The method of event selection varies widely among researchers. The high variability of Fds and the different methodical approaches are responsible for the disparity between the two event catalogs [e.g. \cite{Marcz1997}]. Besides the fully automated approach developed by \cite{Okike2019}, other existing Fd lists \citep[e.g.][]{Oh2008,light:2020,Dumbovic:2024} are created by the manual/sem-automated techniques. In this case study approach, the old and traditional method of extracting the Fd event part from the onset to the Fd minimum is adopted. The magnitude for each of the events is then calculated using Equation \ref{equ1}.

\begin{equation}
\label{equ1}
Fd (\%) = (\frac{|I_{i}-I_{m}|}{I_{m}})100\%,
\end{equation}
where $I_{i}$ is the daily or hourly CR intensity, $I_{m}$ is an average of the intensity variation.
The value of $I_{m}$ may be static or variable, depending on the investigator's choice. The static mean is the value of $I_{m}$ obtained when averaging the data over the entire period. Furthermore, the absence of a time factor in the Equation means that researchers must calculate the magnitude and event time separately, leading to results that depend on the researcher.

A holistic approach is adopted in the current work. Figure \ref{Figureraw} displays the CR intensity variation for the period understudy for the five stations. Rather than extracting part of each of the events, normalizing and expressing the magnitude of each as a percentage, Equation \ref{equ1} is applied to the whole dataset. This expresses the flux variations as a percentage of the mean variation for the period under investigation (1998-2006). The down spikes are indicators of Fds. The giant Fd event of 31/10/2003 is outstanding. Other large events such as those of 19th January 2005 and 15 December 2006 (the first large event from the right) are also evident in the diagram. While some of the large events can easily be spotted by the naked eye, the numerous small Fds are much more difficult to handle. The influence of the solar cycle is also clearly evident in the diagram. It is also important to observe that the SOPO station seems to register higher intensity variations. 

A careful inspection of the diagram suggests that, on average, the flux variations at SOPO stations are comparatively larger than those at other stations. One of the major drawbacks of the analysis of Fds is the difficult task of identifying the magnitude and timing of each of the depressions reflected in Figure \ref{Figureraw} for the different stations. Rather than calculating the magnitudes and time of occurrence of these dips by the visual methods, different versions of the code developed by \citep{Okike2019} were employed by \citet{Okike2020a}, \citet{Okike2020b}, \citet{OkikeNwuzor2020} and \citet{Alhassan2022b} to do the job. The influence of solar cycle oscillations on the magnitude and timing of Fds was not accounted for in these works. Figure 3-5 of \citet{Alhassan2022b} manifests the solar cycle oscillation as suggested by Figure \ref{Figureraw}.

We can calculate event magnitude and timing directly from Figure \ref{Figureraw}. However, in the current work, the contribution from the solar cycle oscillation will be removed before calculating the event magnitude and timing. For the first time, \citet{ok:2021} extended the method of Big Data analysis to the analysis of Fds measured at Climax station. Large volumes of data covering 54 years were passed through a high-pass filter to remove the influence of solar cycles. We will pursue the same method in the current work. 

Figure \ref{Figure2} presents a transformed form of the data in Figure \ref{Figureraw}. A high-pass filter is used to remove the Sun's influence from the raw CR data. The result is presented in Figure \ref{Figure2}. A comparison of Figures \ref{Figure2} and \ref{Figureraw} shows that the observed solar oscillation in Figure \ref{Figureraw} has been disentangled in Figure \ref{Figure2} for each of the stations. It is also important to observe that there are significant intensity variations between some of the stations in the two diagrams. Take SOPO and CALG, for example. In Figure \ref{Figureraw}, SOPO registers higher intensity variations in many of the large and moderate depressions. But we seem to have a completely different scenario in Figure \ref{Figure2}. Here, CALG registers significantly higher intensity variations than SOPO and the rest of the stations. This applies to almost all events. For the large event of 31/10/2003, for example, it can be visually inferred that the magnitude is almost twice at CALG of its value at other stations. 

Data for each of the stations in Figure \ref{Figure2} is {expressed in percentage and passed onto the Fd-Location code. Figure \ref{Figure3} is the result for the CALG station. The blue and red empty small circles stand for small ($Magnitude >-3\%$) and large ($Magnitude <-3\%$) Fds. Besides marking the events on the diagram, the algorithm automatically tabulates the magnitude and the time of maximal depression. The result for one more station is presented for completeness. Figure \ref{Figure4} is for the OULU station. A comparison of Figures \ref{Figure3} and \ref{Figure4} shows that the event magnitude at the two stations varies significantly. {Four events that are simultaneous at the two stations are marked with green-filled cycles. From left to right, the dates of these events are 16/7/2000, 31/10/2003, 19/1/2005 and 13/9/2005. The magnitude of the events at CALG and OULU are respectively -46.52 and -13.93; -75.81 and -22.94; -56.93 and -18.05; and -49.71 and -15.28. This is an indication that CR intensity variation is greater at CALG than at OULU during the period of these FDs. These significant differences in event magnitude are attractive and deserve further inquiry.

The first is to consider the characteristics of the two NMs. The rigidity of OULU (0.77) is less than that of CALG (1.08 GV) (see Table \ref{tb:NM_station}). But this cannot explain the differences as OULU should rather measure higher intensity variations given the low rigidity. Although the difference between the latitudes may play a role, the key factor that may account for the higher intensity variation at CALG may be the altitude. \citet{Pyle1997} investigated the sensitivity of Climax ($R_{c}$ = 3.0 GV, Alt = 3400 M) and Deep River ($R_{c}$ = 1.0 GV, Alt = 145 M); and those of South Pole ($R_{c}$ = 0.09 GV, Alt = 2820 M) and McMurdo ($R_{c}$ = 0.00 GV, Alt = 48 M) NMs using two FD events and the normalized counting rates of the NMs. They established that intensity variation at stations at higher altitudes is higher, the higher geomagnetic vertical cutoff rigidity of such stations notwithstanding. Intensity variation at Climax was 16\% greater than Deep River during the FD event of 2/26/1992, for instance. Rather than the case event approach employed by \citet{Pyle1997},  \citet{OkikeNwuzor2020} used a large number of FD events to re-examine the factors that determine CR intensity variations (sensitivity) seen by NMs.

Besides altitude and rigidity, \citet{OkikeNwuzor2020} concluded that there were several other potential factors like detector size, meteorological conditions, and features of operation that may influence CR variations measured at each station. A comparison of the CR signals presented in Figures \ref{Figureraw} and \ref{Figure2} suggests that the 11-year solar oscillation (refer to Section \ref{Analysis}) $-$ which contributes to the major differences between the time-intensity profile of the signals in the two Figures $-$ is another factor that may impact significantly on the flux variations at different NMs. For the purpose of comparison, the four events highlighted with green colors in Figures \ref{Figure3} and \ref{Figure4} are clearly labeled with dates in Figures \ref{Figureraw} and \ref{Figure2}. Figure \ref{Figureraw} shows that SOPO sees the greatest intensity depressions for the events of 16/7/2000, 31/10/2003, and 13/9/2005. There are also many other reductions that suggest that SOPO sees larger intensity variations. However, CALG measures the largest variation for the event of 19/1/2005. It is also easy to point to some other flux reductions where CALG and other stations (e.g. NVBK) see larger depressions than SOPO in this diagram. But we have a different picture in Figure \ref{Figure2}. For each of the four events and all other depressions, CALG consistently registers larger depressions. Further, the intensity changes in the two Figures during the events are quite significant. For example, the observed differences between the size of reductions for the events of 16/7/2000 and 31/10/2003 in Figure \ref{Figureraw} seem to double in Figure \ref{Figure2}.

To further illustrate the role impact of the long-term solar influence of the Sun on the amplitude of FDs, the large event that happened within the period of solar minimum of Cycle 23 (event of 15/12/2006) is indicated in Figures \ref{Figureraw}-\ref{Figure4}. While the event is labeled ``15/12/2006" in Figures \ref{Figure2}-\ref{Figure4}, the small size of it in Figure \ref{Figureraw} warrants that an arrow (blue) is used to point to it. The large difference in the magnitude of this event at the four stations (SOPO has no data for the period) before and after removing the 11-year solar cycle influence is evidence that the superposition tendencies of CR data might have serious implications for small FDs as recently submitted by \citet{ok:2024}. The overall intensity variations in Figures \ref{Figureraw} and \ref{Figure2} suggest that the influence of the 11-year cycle may vary between locations. The relatively large amplitudes of events at CALG station in Figure \ref{Figure2} suggests the need to investigate the impact of solar cycle variation at these stations in detail. However, we leave such investigation for future work.

Given the significant disparity between intensity variations at CALG and other stations, there is a need to test the consistency of event magnitude and timing at CALG with those at another NM. FDs at three other stations including OULU, APTY and SOPO were used to validate the FDs at CALG. The correlation coefficient (r) between simultaneous FDs at CALG and OULU is 0.981 (N = 177); r for CALG and APTY is 0.977 (173); r for CALG and SOPO is 0.984 (N = 153). Note that N is the number of simultaneous FDs at pairs of stations.}

\begin{figure}
	\includegraphics[width=\columnwidth]{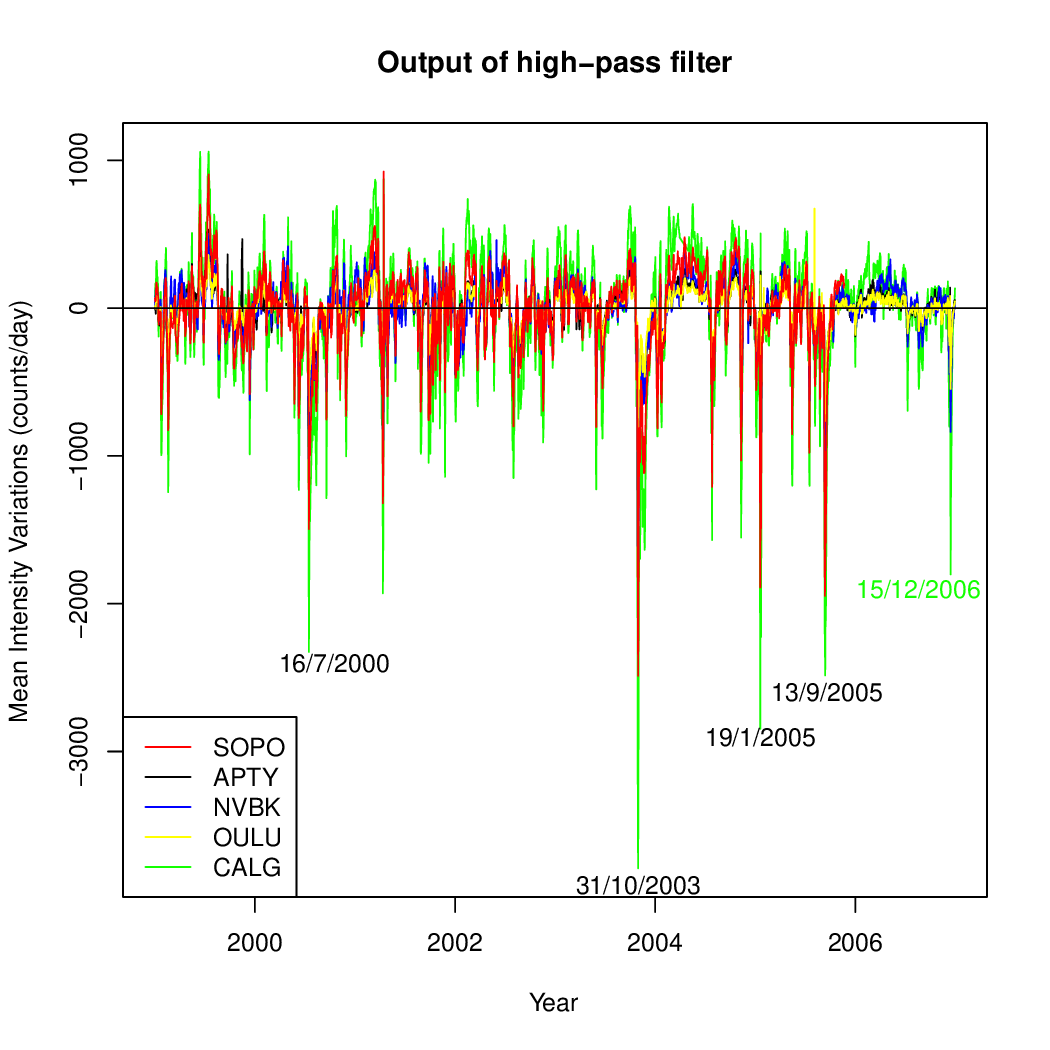}
    \caption{The output of the high-pass filter (used to remove the contribution from the 11-year oscillations at each of the five stations). The input is raw daily CR data.}
    \label{Figure2}
\end{figure}

\begin{figure}
	\includegraphics[width=\columnwidth]{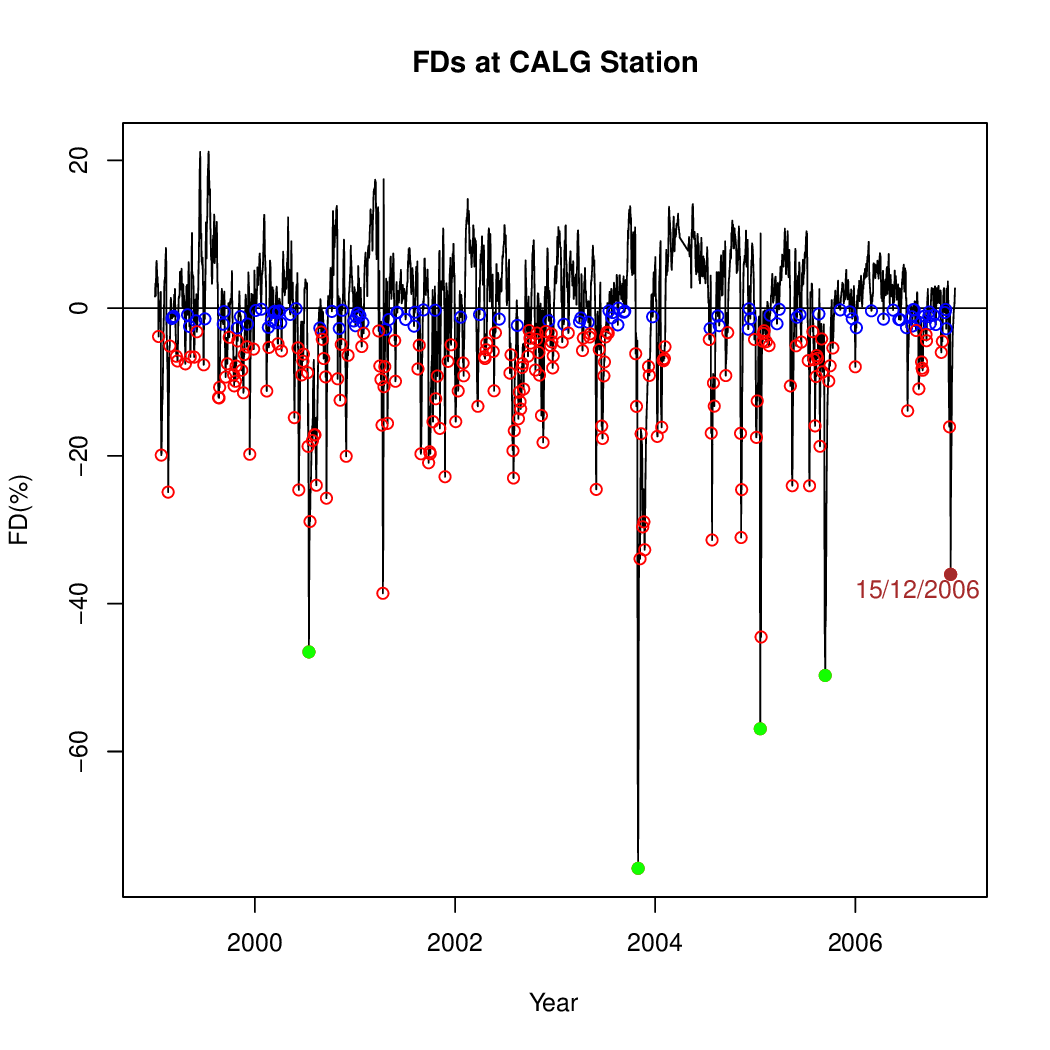}
    \caption{Plot of Fd event magnitude and timing at CALG station. The blue and red empty small circles stand for small ($Magnitude >-3\%$) and large ($Magnitude <-3\%$) Fds. Some of the FDs that are simultaneous at CALG and OULU stations are marked with green filled colors.}
    \label{Figure3}
\end{figure}

\begin{figure}
	\includegraphics[width=\columnwidth]{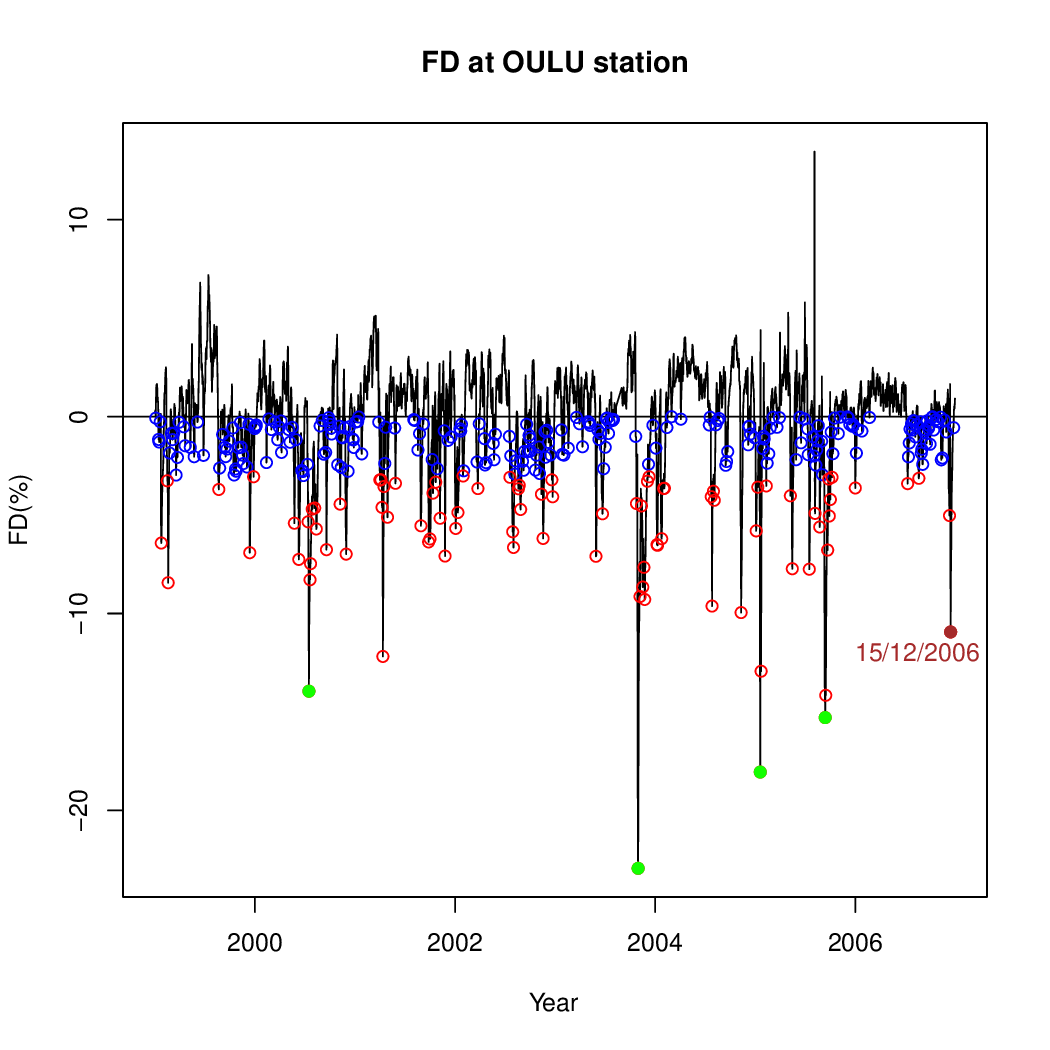}
    \caption{Same as Figure \ref{Figure3} but for OULU station.}
    \label{Figure4}
\end{figure}

 \section{Results and Discussion}
\subsection{Fds and their Relations with Simultaneous interplanetary parameters and geomagnetic activity indices data at individual NMs}\label{Individual}
{The number of Fds at APTY, CALG, NVBK, OULU, and SOPO are respectively 317(2922), 302(2886), 341(2922), 324(2901) and 274(2510). The number in brackets indicates the days each of the stations had an observation. The number of days between 1/1/1999 and 31/12/2006 is 2922, implying that only APTY and NVBK had complete observation within the period. It is, thus, expected that the two stations should see a relatively larger number of Fds than the rest of the stations. While this is true for NVBK, it does not hold for APTY. The number of Fds measured at OULU is greater than those at APTY. This suggests that other factors account for the number of Fds measured at a particular location. Solar-terrestrial variables are some of the parameters that may have a relationship between Fds at different locations. 

The connection between Fds at the APTY station and IMF, SWS, Kp, Dst, and Ap are tested. This is presented in Table \ref{APTYFD}. The results presented here are quite insightful. Besides the absolute values and poor correlation coefficients, r, commonly reported in the literature \citep[e.g.][]{Belov2001, kane:2010}, the presented signs for the relations are physically meaningful. While \citet{kane:2010}, for instance, found zero correlation for a combination of large and small Fds, they reported a negative correlation of -0.7 for the six largest Fds in their catalog. For the 17 events in their list, they find a weak but positive correlation between Dst and Fd. On the contrary, \citet{Badruddin:2015} reported a strong positive r of 0.58 between Fd and SWS. \citet{Kharayat:2016} found a positive correlation of 0.79 between Dst and CR intensity. \citet{Lingri:2016} find only a marginal correlation of 0.02 between Fds and Dst. In light of these contradictory reports on the connection between Fds and solar-terrestrial variables (see also \citet{Ghag:2023}) and the references therein, the result presented in Figure \ref{APTYFD} is worth more attention.

Inspection of Table \ref{APTYFD} shows that IMF, SWS, Kp, and Ap all register negative correlations between Fds. These negative relations are interestingly replicated for both the strong and weak Fds. All the correlations, except the two values marked with `*', reported in the Table are statistically significant at levels between 90 to 99.8\%. The relation between Fd and IMF, Fd and Ap are not significant at 90\% level.

It is also interesting to observe that the negative relations between Fds and IMF, SWS, Kp, and Ap also hold for Dst and IMF, SWS, Kp, and Ap. These consistent negative relations between Fds, Dst and other solar-geomagnetic variables are suggestive of the fact that the time evolution of  Fds and Dst may be quite similar \citep[see][and the references within]{Ghag:2023}. This contradicts the submission of \citet{kane:2010}. We also observe that almost all the corresponding relations for Dst are higher than those Fds. For example, the r for Dst and IMF in the top panel of Table \ref{APTYFD} is -0.56 whereas it is -0.37 for Fd and IMF. While the negative signs are an indication that Fd and Dst are caused by similar space weather conditions, the larger the values of r's associated with Dst could be a pointer to the delay between the two events. \citet{Ghag:2023} and the references therein suggest that storms lag behind Fds for a period of about 3-4 hours. Though \citet{Mishra:2022} did not find any delay between Fd and Dst, the earlier response of CRs to common causing agents (CMEs/ICMEs) is expected since the exclusion of CR particles by the solar ejections/shocks happened in remote distant places in the IP medium well before the solar phenomena arrive the Earth to interact with the Earth's magnetic field. However, the smaller r (especially for small Fds) obtained for Fds between these variables implies that other factors interfere with the interactions between Fds and their simultaneous solar sources. The smallest r's are obtained for small Fds, suggesting more pronounced effects of such agents on small amplitude events. The recent work of \citet{ok:2024} empirically demonstrated that CR anisotropy and the 11-year cycle oscillations are the major factors that exert significant influence on Fds.  Of course, Dst does not exhibit the 11-year oscillation observed in CR data. They do not also suffer contamination from diurnal variations.

For validation purposes, correlation coefficients for Fds measured at another station (NVBK) are presented in Table \ref{NVBKFD}. All the negative and positive r's between Fds and other variables are repeated in the Table. The strong and highly statistically significant r's are also captured. The smallest r's are also registered for weak Fds in bottom panel of Table \ref{NVBKFD}. While small Fds have a significant relation with SWS, Kp and Dst at APTY, none of the five variables has a significant connection with Fd at NVBK NM.

\begin{table}
\caption{The correlation coefficients, r, between Fds at APTY and solar-terrestrial parameters are presented.}
\label{APTYFD}
\begin{tabular}{llrrrr}
  \hline
All & IMF & SWS & Kp & Dst & Ap \\ 
  \hline
Fd    & -0.37 & -0.47 & -0.38 & 0.47 & -0.43 \\ 
  IMF    &   & 0.22 & 0.62 & -0.56 & 0.67 \\ 
  SWS  &   &   & 0.61 & -0.45 & 0.49 \\ 
  Kp   &   &   &   & -0.73 & 0.87 \\ 
  Dst    &   &   &   &   & -0.79 \\ 

  \hline
Big   &  &  &  &  &  \\ 
  \hline
Fd &    -0.29 & -0.61 & -0.42 & 0.47 & -0.45 \\ 
  IMF &      & 0.32 & 0.58 & -0.56 & 0.64 \\ 
  SWS &      &   & 0.66 & -0.53 & 0.59 \\ 
  Kp &   &      &   & -0.72 & 0.90 \\ 
  Dst &  &      &   &   & -0.76 \\ 

  \hline
Small &    &  &  &  &  \\ 
  \hline
Fd &   -0.10* & -0.14 & -0.13 & 0.12 & -0.11* \\ 
  IMF &     & -0.01 & 0.61 & -0.45 & 0.66 \\ 
  SWS &      &  & 0.52 & -0.27 & 0.31 \\ 
  Kp &   &      &  & -0.72 & 0.87 \\ 
  Dst &   &      &   &  & -0.79 \\ 
   \hline
\end{tabular}
\begin{small}
 Note:  'All', `Big', and `Small' respectively stand for all the 317 Fds (large and small inclusive), large Fds (113), and small Fds (204) at APTY station. Numbers marked with `*' are not statistically significant at 90\%.
 \end{small}
\end{table}

\begin{table}
\caption{The correlation coefficients, r, between Fds at NVBK and solar-terrestrial parameters are presented.}
\label{NVBKFD}
\begin{tabular}{lrrrrr}
  \hline
All  & IMF & SWS & Kp & Dst & Ap \\ 
  \hline
Fd &    -0.35 & -0.44 & -0.37 & 0.47 & -0.42 \\ 
  IMF &      & 0.21 & 0.60 & -0.53 & 0.66 \\ 
  SWS &      &   & 0.64 & -0.48 & 0.51 \\ 
  Kp &   &      &   & -0.72 & 0.87 \\ 
  Dst &   &      &   &   & -0.77 \\ 
 
  \hline
Big    &  &  &  &  &  \\ 
  \hline
Fd &    -0.31 & -0.60 & -0.43 & 0.47 & -0.46 \\ 
  IMF &      & 0.32 & 0.60 & -0.52 & 0.66 \\ 
  SWS &      &   & 0.67 & -0.57 & 0.58 \\ 
  Kp &   &     &   & -0.72 & 0.88 \\ 
  Dst &   &      &   &   & -0.74 \\ 
  
  \hline
Small   &  &  &  &  &  \\ 
  \hline
Fd &    -0.09* & -0.10* & -0.07* & 0.07* & -0.06* \\ 
  IMF &      & -0.02 & 0.57 & -0.43 & 0.62 \\ 
  SWS &     &   & 0.57 & -0.31 & 0.38 \\ 
  Kp &      &   &   & -0.71 & 0.88 \\ 
  Dst &      &   &   &   & -0.79 \\ 
   \hline
\end{tabular}
\begin{small}
Note:`All', `Big', and `Small' respectively stand for all the 341 Fds (large and small inclusive), large Fds (150), and small Fds (191) at APTY station. Numbers marked with `*' are not statistically significant at 90\%. 
\end{small}
\end{table}



\subsection{Fds and their Relations with Simultaneous interplanetary parameters and geomagnetic activity indices data at two NMs}
 The relation tested in Section \ref{Individual} can be extended to two NMs. The number of simultaneous Fds at the following pairs of stations APTY/ CALG, APTY/NVBK, APTY/ OULU, APTY/ SOPO, CALG/ NVBK, CALG /OULU, CALG/ SOPO, NVBK/ OULU, NVBK/ SOPO, and OULU /SOPO are respectively 173, 169, 236, 142, 152, 177, 153, 169, 137, and 149. The results are presented in Table \ref{two}.

The r's between the simultaneous Fds at the four pairs of stations are very high (r $\approx 1$) [Table (\ref{two}) Column 2]. These high correlation coefficients are also reflected in the r between Fds at the two stations and other variables. The r between Fd and IMF at APTY and CALG stations are respectively -0.35 and -0.39. For SWS, they are -0.56 and -0.54. The close r's are a pointer that two carefully selected Fd catalogues would yield similar results for strong/simultaneous Fds irrespective of the location of the NM. The results presented here reflect those of large Fds in Tables \ref{APTYFD} and \ref{NVBKFD}.}
\begin{table}
\caption{The correlation coefficients, r, between simultaneous Fds at APTY and CALG, NVBK, OULU, SOPO with IMF, SWS, Kp, Dst,  and Ap 
.}
\label{two}
\begin{tabular}{lrrrrrr}
  \hline
 & FD$_{ca}$ & IMF & SWS & Kp & Dst & Ap \\ 
  \hline
FD$_{ap}$   & 0.98 & -0.35 & -0.56 & -0.38 & 0.47 & -0.43 \\ 
  FD$_{ca}$    &   & -0.39 & -0.54 & -0.38 & 0.49 & -0.44 \\ 
  IMF    &   &   & 0.21 & 0.62 & -0.58 & 0.68 \\ 
  SWS    &   &   &   & 0.55 & -0.44 & 0.48 \\ 
  Kp   &   &   &   &   & -0.75 & 0.89 \\ 
  Dst    &   &   &   &   &   & -0.77 \\ 

  \hline
  & FD$_{nv}$ & IMF & SWS & Kp & Dst & Ap \\
  \hline
FD$_{ap}$   & 0.96 & -0.35 & -0.54 & -0.37 & 0.47 & -0.41 \\ 
  FD$_{nv}$    &   & -0.33 & -0.53 & -0.34 & 0.43 & -0.39 \\ 
  IMF    &   &   & 0.24 & 0.64 & -0.61 & 0.70 \\ 
  SWS    &   &   &   & 0.55 & -0.44 & 0.46 \\ 
  Kp    &   &   &   &   & -0.74 & 0.89 \\ 
  Dst    &   &   &   &   &   & -0.77 \\ 

  \hline
 & FD$_{ou}$ & IMF & SWS & Kp & Dst & Ap \\
  \hline
FD$_{ap}$    & 0.98 & -0.38 & -0.51 & -0.39 & 0.46 & -0.44 \\ 
  FD$_{ou}$    &   & -0.38 & -0.49 & -0.38 & 0.46 & -0.44 \\ 
  IMF    &   &   & 0.24 & 0.64 & -0.57 & 0.68 \\ 
  SWS    &   &   &   & 0.58 & -0.44 & 0.49 \\ 
  Kp    &   &   &   &   & -0.73 & 0.88 \\ 
  Dst    &   &   &   &   &   & -0.78 \\ 
 
  \hline
  & FD$_{so}$ & IMF & SWS & Kp & Dst & Ap \\ 
  \hline
FD$_{ap}$    & 0.96 & -0.32 & -0.52 & -0.29 & 0.39 & -0.36 \\ 
  FD$_{so}$    &   & -0.37 & -0.56 & -0.39 & 0.49 & -0.45 \\ 
  IMF    &   &   & 0.17 & 0.64 & -0.58 & 0.69 \\ 
  SWS    &   &   &   & 0.47 & -0.39 & 0.41 \\ 
  Kp     &   &   &   &   & -0.74 & 0.90 \\ 
  Dst    &   &   &   &   &   & -0.76 \\ 
   \hline
\end{tabular}
\begin{small}
Note: Number of simultaneous Fds at APTY and CALG = 173,  APTY and NVBK, = 169,  APTY and OULU = 236,  NVBK and SOPO = 137(small and big events inclusive).
\end{small}
\end{table}

\subsection{Fds and their relations with Simultaneous interplanetary parameters and geomagnetic activity indices data at the five NMs}
The simultaneous Fds and their associated interplanetary parameters and geomagnetic activity indices are presented in Table \ref{Tab:Fd}. The Fds that are simultaneously detected at the five stations can be taken as strong or large events. There are eighty in number. While it may be easy to define event magnitude as large($<-3\%$) or small($>-3\%$) with reference to a particular NM, Table \ref{Tab:Fd} shows that it is not easy to do so with Fd data from an array of CR stations. The magnitude of the first event (14/1/1999) in the Table varies appreciably across the five stations with only CALG measuring a magnitude $\leq -3\%$. Others are much smaller. The second event (18/2/1999) is obviously large. But its magnitude equally varies significantly among the five stations with CALG also measuring the largest magnitude. The third event (10/3/1999) all pass the common test of (CR(\%) $\leq -3$). Nevertheless, the intensity variation changes. This time, SOPO measures the highest flux variation. 

An over of the event magnitude in the Table shows that CALG measures larger event magnitudes, especially for relatively large events. The magnitudes of some relatively large Fds in the Table are 2 or more times their values at CALG. The second event, for instance, has a magnitude of 8.44 at OULU whereas its magnitude at CALG is 24.9\% (about 3 times). The same applies to some other very large Fds, including those of 9/6/2000, 16/7/2000, 31/10/2003, and 19/1/2005. The observed large differences are reflected in Figures \ref{Figureraw} and \ref{Figure2} and can be a reflection of the characterisitics of the NMs as suggested in Table \ref{tb:NM_station}. The latitude, longitude, altitude, rigidity and other local dependen CR phenomena could be the reason why CALG measures larger Fd event magnitude.

The r's for the five stations and the associated interplanetary parameters and geomagnetic activity indices are presented in Table \ref{five}. The pattern of relationship observed in Tables \ref{APTYFD}-\ref{two} are repeated here.

\begin{table*} 
\caption{Simultaneous Forbush deceases, Interplanetary  parameters and geomagnetic indices}
\label{Tab:Fd}
\begin{tabular*}{\textwidth}{@{\extracolsep\fill}cccccccccccc} 
\hline
S/N &Date	&	IMF	&	SWS	&	kp	&	Dst	&	ap &	FD$_{ou}$	&	FD$_{so}$	&	FD$_{ap}$	&	FD$_{ca}$	&	FD$_{nv}$ \\
 & &	(nT)&	(kms$ ^{-1}) $&		&		(nT)&	 (nT)	&	(\%)	&	(\%)	&	(\%)	&	(\%)	&	(\%) \\
\hline
1 & 1999-01-14 & 12.20 & 461 &  40 & -67 &  29 & -1.16 & -2.22 & -1.86 & -3.84 & -0.16 \\ 
  2 & 1999-02-18 & 17.10 & 599 &  60 & -84 &  80 & -8.44 & -16.52 & -11.65 & -24.90 & -10.20 \\ 
  3 & 1999-03-10 & 7.80 & 441 &  40 & -44 &  34 & -0.84 & -2.13 & -0.11 & -1.12 & -0.17 \\ 
  4 & 1999-09-16 & 6.20 & 572 &  40 & -46 &  31 & -2.03 & -4.62 & -2.88 & -9.33 & -4.80 \\ 
  5 & 1999-09-29 & 6.80 & 539 &  37 & -30 &  25 & -1.25 & -2.07 & -3.03 & -3.95 & -3.18 \\ 
  6 & 1999-12-13 & 11.40 & 489 &  33 & -46 &  26 & -6.91 & -11.64 & -8.43 & -19.78 & -12.45 \\ 
  7 & 2000-05-24 & 13.70 & 636 &  60 & -90 &  93 & -5.41 & -12.95 & -7.01 & -14.82 & -2.80 \\ 
  8 & 2000-06-09 & 10.20 & 609 &  13 & -34 &   5 & -7.25 & -14.88 & -7.94 & -24.60 & -13.56 \\ 
  9 & 2000-06-20 & 6.20 & 379 &  17 &   5 &   6 & -2.74 & -6.78 & -3.02 & -9.03 & -4.98 \\ 
  10 & 2000-06-24 & 9.00 & 551 &  27 & -20 &  15 & -2.79 & -5.82 & -3.54 & -7.19 & -5.29 \\ 
  11 & 2000-06-26 & 11.50 & 512 &  43 & -36 &  40 & -2.99 & -5.40 & -3.30 & -6.24 & -3.64 \\ 
  12 & 2000-07-11 & 13.70 & 458 &  43 &  13 &  34 & -2.43 & -5.96 & -3.12 & -8.74 & -2.71 \\ 
  13 & 2000-07-16 & 21.80 & 816 &  43 & -172 &  50 & -13.94 & -29.89 & -16.91 & -46.52 & -23.33 \\ 
  14 & 2000-08-12 & 25.00 & 599 &  67 & -128 & 123 & -5.71 & -13.95 & -7.35 & -23.97 & -10.17 \\ 
  15 & 2000-09-18 & 19.20 & 744 &  53 & -103 &  70 & -6.76 & -15.07 & -8.15 & -25.72 & -11.28 \\ 
  16 & 2000-10-29 & 13.70 & 381 &  40 & -89 &  34 & -2.45 & -6.12 & -2.32 & -9.56 & -4.07 \\ 
  17 & 2000-11-07 & 20.20 & 512 &  43 & -89 &  46 & -4.45 & -11.03 & -4.85 & -12.47 & -4.16 \\ 
  18 & 2000-11-11 & 7.20 & 804 &  30 & -35 &  16 & -2.58 & -5.77 & -2.82 & -4.91 & -2.55 \\ 
  19 & 2000-11-29 & 9.20 & 512 &  47 & -81 &  56 & -6.98 & -14.57 & -8.63 & -20.06 & -9.75 \\ 
  20 & 2000-12-27 & 7.70 & 390 &  20 &  -1 &   8 & -1.57 & -1.99 & -1.31 & -1.73 & -1.47 \\ 
  21 & 2001-04-01 & 7.50 & 746 &  43 & -137 &  38 & -3.21 & -5.52 & -2.86 & -7.83 & -3.17 \\ 
  22 & 2001-04-05 & 7.50 & 617 &  33 & -31 &  19 & -3.21 & -6.08 & -2.94 & -9.66 & -5.68 \\ 
  23 & 2001-04-09 & 8.60 & 622 &  33 & -53 &  20 & -4.61 & -11.68 & -5.41 & -15.83 & -8.13 \\ 
  24 & 2001-04-12 & 15.10 & 659 &  40 & -131 &  50 & -12.18 & -26.39 & -14.34 & -38.60 & -20.93 \\ 
  25 & 2001-04-19 & 7.90 & 436 &  17 & -41 &   6 & -2.38 & -4.65 & -2.77 & -7.92 & -4.89 \\ 
  26 & 2001-04-22 & 11.80 & 360 &  43 & -55 &  37 & -0.54 & -3.62 & -0.52 & -2.88 & -1.65 \\ 
  27 & 2001-04-29 & 7.60 & 596 &  23 & -18 &  13 & -5.11 & -11.93 & -6.45 & -15.60 & -9.48 \\ 
  28 & 2001-05-28 & 9.10 & 505 &  33 &  -8 &  18 & -3.39 & -6.46 & -4.67 & -9.90 & -7.43 \\ 
  29 & 2001-08-18 & 11.80 & 518 &  27 & -43 &  15 & -1.70 & -5.82 & -2.71 & -8.23 & -1.84 \\ 
  30 & 2001-09-26 & 10.70 & 549 &  33 & -72 &  26 & -6.37 & -15.15 & -7.99 & -20.93 & -8.49 \\ 
  31 & 2001-10-22 & 15.10 & 578 &  60 & -150 &  96 & -3.33 & -8.37 & -5.22 & -12.27 & -3.10 \\ 
  32 & 2001-10-28 & 11.20 & 450 &  47 & -99 &  44 & -2.65 & -4.79 & -3.13 & -9.19 & -2.95 \\ 
  33 & 2001-11-25 & 11.50 & 650 &  20 & -106 &   8 & -7.08 & -11.40 & -8.59 & -22.82 & -8.23 \\ 
  34 & 2001-12-17 & 8.80 & 471 &  30 & -30 &  16 & -1.04 & -3.01 & -1.26 & -4.97 & -1.34 \\ 
  35 & 2002-01-03 & 5.90 & 342 &   7 & -16 &   2 & -5.68 & -9.44 & -6.30 & -15.36 & -8.89 \\ 
  36 & 2002-02-01 & 11.20 & 347 &  27 & -17 &  14 & -2.75 & -4.70 & -1.97 & -9.16 & -5.68 \\ 
  37 & 2002-04-20 & 10.10 & 563 &  53 & -106 &  70 & -2.46 & -6.95 & -3.25 & -6.83 & -5.00 \\ 
  38 & 2002-04-24 & 7.10 & 488 &  17 & -30 &   7 & -2.34 & -5.22 & -2.96 & -5.63 & -4.93 \\ 
  39 & 2002-05-23 & 17.00 & 606 &  47 & -38 &  78 & -2.19 & -7.31 & -2.29 & -11.19 & -4.52 \\ 
  40 & 2002-07-20 & 7.40 & 789 &  30 & -20 &  18 & -3.08 & -5.21 & -4.07 & -8.80 & -4.44 \\ 
  41 & 2002-08-20 & 7.20 & 479 &  33 & -48 &  30 & -3.66 & -9.38 & -4.44 & -14.97 & -7.86 \\ 
  42 & 2002-08-28 & 8.90 & 447 &  17 & -19 &   7 & -4.72 & -8.12 & -5.42 & -13.65 & -7.32 \\ 
  43 & 2002-09-08 & 11.70 & 479 &  33 & -101 &  36 & -2.72 & -6.64 & -3.01 & -10.95 & -3.89 \\ 
  44 & 2002-10-01 & 19.50 & 388 &  50 & -100 &  67 & -1.11 & -1.43 & -0.38 & -4.08 & -1.46 \\ 
  45 & 2002-10-03 & 11.50 & 464 &  47 & -78 &  45 & -1.75 & -1.86 & -1.71 & -4.74 & -2.12 \\ 
  46 & 2002-10-25 & 6.80 & 689 &  43 & -68 &  39 & -1.96 & -2.76 & -1.98 & -4.24 & -1.58 \\ 
  47 & 2002-11-03 & 9.70 & 478 &  43 & -65 &  35 & -1.55 & -3.34 & -2.16 & -6.01 & -1.82 \\ 
  48 & 2002-11-12 & 12.40 & 569 &  30 & -15 &  17 & -3.94 & -5.62 & -5.25 & -14.53 & -6.20 \\ 
  49 & 2002-12-20 & 6.10 & 528 &  33 & -47 &  21 & -3.21 & -4.28 & -2.58 & -4.58 & -3.39 \\ 
  50 & 2002-12-23 & 10.10 & 517 &  37 & -42 &  24 & -4.07 & -7.05 & -4.06 & -8.09 & -6.43 \\ 

\hline
\end{tabular*}
Note: FD$_{ou}$ denotes Forbush decreases from OULU neutron monitor station, FD$_{so}$ denotes Forbush decreases from SOPO neutron monitor station, FD$_{ap}$ denotes Forbush decreases from APTY neutron monitor station, FD$_{ca}$ denotes Forbush decreases from CALG neutron monitor station and FD$_{nv}$ denotes Forbush decreases from NVBK neutron monitor station.
\end{table*}

\begin{table*} 
\label*{Table 5 cont}\\
\begin{tabular*}{\textwidth}{@{\extracolsep\fill}cccccccccccc} 
\hline
S/N &Date	&	IMF	&	SWS	&	kp	&	Dst	&	ap &	FD$_{ou}$	&	FD$_{so}$	&	FD$_{ap}$	&	FD$_{ca}$	&	FD$_{nv}$ \\
 & &	(nT)&	(kms$ ^{-1}) $&		&		(nT)&	 (nT)	&	(\%)	&	(\%)	&	(\%)	&	(\%)	&	(\%) \\
\hline

51 & 2003-01-27 & 8.70 & 507 &  20 &  -5 &  10 & -1.92 & -3.98 & -3.16 & -4.60 & -1.58 \\ 
  52 & 2003-05-02 & 4.70 & 583 &  27 & -24 &  15 & -0.26 & -1.90 & -1.20 & -1.92 & -1.13 \\ 
  53 & 2003-05-09 & 8.30 & 791 &  40 & -27 &  31 & -0.45 & -2.70 & -0.77 & -3.48 & -1.91 \\ 
  54 & 2003-06-23 & 7.50 & 502 &  33 & -20 &  18 & -4.94 & -10.87 & -5.29 & -17.64 & -8.87 \\ 
  55 & 2003-10-25 & 15.40 & 540 &  30 & -26 &  16 & -4.41 & -9.69 & -5.69 & -13.30 & -7.28 \\ 
  56 & 2003-10-31 & 15.80 & 1003 &  63 & -117 & 116 & -22.94 & -49.76 & -28.35 & -75.81 & -36.13 \\ 
  57 & 2003-11-07 & 5.80 & 509 &  20 &  -9 &   8 & -9.14 & -20.83 & -12.06 & -33.92 & -16.19 \\ 
  58 & 2003-11-21 & 9.60 & 513 &  40 & -140 &  42 & -7.65 & -22.32 & -10.14 & -28.92 & -9.84 \\ 
  59 & 2003-12-08 & 11.10 & 519 &  43 & -32 &  35 & -2.42 & -7.55 & -4.27 & -7.90 & -4.89 \\ 
  60 & 2004-01-25 & 9.90 & 472 &  43 & -65 &  38 & -6.20 & -12.77 & -7.13 & -16.13 & -9.38 \\ 
  61 & 2004-02-01 & 7.50 & 533 &  23 & -11 &  10 & -3.65 & -5.42 & -4.78 & -7.13 & -4.06 \\ 
  62 & 2004-07-24 & 16.90 & 561 &  43 & -13 &  37 & -4.07 & -10.55 & -5.05 & -16.91 & -7.60 \\ 
  63 & 2004-07-27 & 17.40 & 904 &  77 & -120 & 186 & -9.62 & -24.39 & -11.02 & -31.38 & -14.53 \\ 
  64 & 2004-08-04 & 5.60 & 334 &   7 & -10 &   3 & -4.25 & -9.97 & -4.78 & -13.27 & -7.01 \\ 
  65 & 2004-09-15 & 4.90 & 549 &  27 & -23 &  14 & -2.47 & -4.59 & -2.87 & -9.13 & -3.95 \\ 
  66 & 2004-09-22 & 7.20 & 477 &  27 & -11 &  16 & -1.78 & -2.33 & -2.53 & -3.31 & -1.95 \\ 
  67 & 2004-11-10 & 18.40 & 691 &  70 & -176 & 161 & -9.95 & -20.82 & -11.25 & -31.05 & -16.30 \\ 
  68 & 2004-12-06 & 9.70 & 424 &  33 & -26 &  20 & -1.43 & -1.85 & -2.11 & -2.85 & -4.75 \\ 
  69 & 2005-01-19 & 12.60 & 840 &  50 & -64 &  60 & -18.05 & -38.00 & -21.75 & -56.93 & -26.26 \\ 
  70 & 2005-01-22 & 13.20 & 766 &  40 & -72 &  33 & -12.93 & -30.01 & -15.63 & -44.50 & -22.69 \\ 
  71 & 2005-02-02 & 6.30 & 507 &  20 & -11 &   9 & -1.13 & -2.68 & -2.49 & -3.06 & -1.56 \\ 
  72 & 2005-03-21 & 7.20 & 444 &  17 & -13 &   8 & -0.54 & -1.33 & -0.12 & -2.12 & -0.35 \\ 
  73 & 2005-05-09 & 8.40 & 620 &  20 & -48 &  10 & -4.02 & -8.19 & -4.93 & -10.53 & -4.94 \\ 
  74 & 2005-05-30 & 15.70 & 469 &  57 & -73 &  90 & -2.19 & -4.85 & -2.41 & -5.10 & -3.47 \\ 
  75 & 2005-07-13 & 6.90 & 560 &  40 & -32 &  31 & -1.93 & -5.87 & -2.81 & -7.11 & -2.89 \\ 
  76 & 2005-07-30 & 5.30 & 507 &  23 & -12 &  10 & -1.02 & -3.11 & -1.40 & -3.18 & -2.68 \\ 
  77 & 2005-08-07 & 5.20 & 657 &  27 & -25 &  14 & -4.92 & -10.54 & -6.03 & -15.94 & -8.17 \\ 
  78 & 2005-08-16 & 8.70 & 615 &  30 &   0 &  14 & -1.23 & -3.46 & -2.03 & -6.79 & -3.85 \\ 
  79 & 2005-08-31 & 13.20 & 388 &  40 & -45 &  49 & -1.25 & -2.93 & -2.08 & -4.17 & -3.16 \\ 
  80 & 2005-09-03 & 6.90 & 596 &  37 & -51 &  27 & -2.95 & -5.40 & -3.71 & -8.63 & -5.31 \\ 
\hline
\end{tabular*}
Note: FD$_{ou}$ denotes Forbush decreases from OULU neutron monitor station, FD$_{so}$ denotes Forbush decreases from SOPO neutron monitor station, FD$_{ap}$ denotes Forbush decreases from APTY neutron monitor station, FD$_{ca}$ denotes Forbush decreases from CALG neutron monitor station and FD$_{nv}$ denotes Forbush decreases from NVBK neutron monitor station.
\end{table*}

\begin{table*}
\caption{Correlation matrix between simultaneous Fds and the associated paramters.}
\label{five}
\centering
\begin{tabular}{rrrrrrrrrr}
  \hline
Variables  & SWS & Kp & Dst & Ap & Fd$_{ou}$ & Fd$_{so}$ & Fd$_{ap}$ & Fd$_{ca}$ & Fd$_{nv}$ \\ 
  \hline
IMF &    0.25 & 0.69 & -0.62 & 0.70 & -0.40 & -0.42 & -0.39 & -0.43 & -0.37 \\ 
  SWS &      & 0.42 & -0.40 & 0.44 & -0.61 & -0.62 & -0.62 & -0.61 & -0.58 \\ 
  Kp &      &   & -0.69 & 0.91 & -0.32 & -0.37 & -0.33 & -0.34 & -0.27 \\ 
  Dst &      &   &   & -0.71 & 0.47 & 0.49 & 0.46 & 0.48 & 0.40 \\ 
  Ap &     &   &   &  & -0.42 & -0.46 & -0.41 & -0.43 & -0.37 \\ 
  Fd$_{ou}$ &      &   &   &   &   & 0.98 & 0.99 & 0.98 & 0.97 \\ 
  Fd$_{so}$ &      &   &   &   &   &   & 0.98 & 0.99 & 0.96 \\ 
  Fd$_{ap}$ &      &   &   &   &   &   &   & 0.98 & 0.96 \\ 
  Fd$_{ca}$ &      &   &   &   &   &   &   &   & 0.97 \\ 
   \hline
\end{tabular}
\end{table*}

\subsection{Simultaneous Forbush decreases -- Interplanetary Parameters and geomagnetic indices association}
To further investigate the level of the interplay of the SFds with interplanetary parameters and solar geomagnetic activity indices, we conducted a two-dimensional regression analysis.
For easy comparison,  plots of the Fds from various NMs against the parameters are made and presented in a composite form as shown in Figure (\ref{fig:Fd_solar_geomagnetic}).
In order to numerically determine the level of association, we use Pearson's correlation at a 5\% significant level.
The correlation coefficients are displayed in the plots (Figure \ref{fig:Fd_solar_geomagnetic}) and are all statistically significant.
Upon analyzing the data across stations, it is evident that the correlation coefficient's magnitude remains consistent.
This consistency in coefficients signifies that the interplanetary and solar geomagnetic activity parameters' effects on CR counts across various stations are comparable, thereby validating that the Fds are simultaneous.
\begin{figure*}[h]
\includegraphics[scale=0.62]{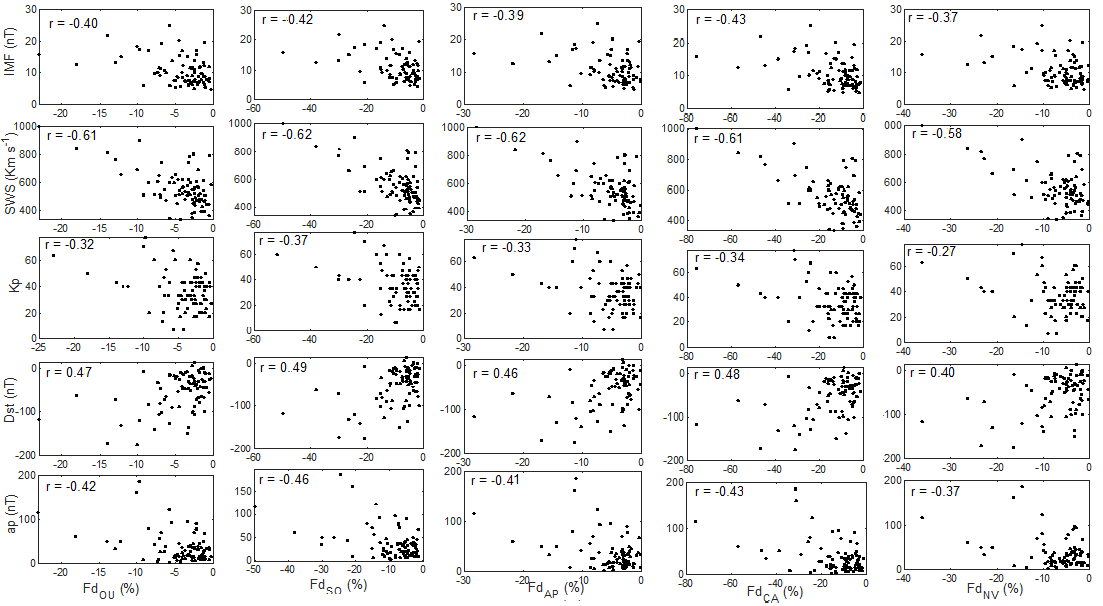}
    \caption{Fd vs interplanetary parameters and geomagnetic activity indices}
    \label{fig:Fd_solar_geomagnetic}
Columns 1 to 5 are for OULU, SOPO, APTY, CALG, and NVBK NMs stations respectively. The mean correlation coefficient $ \bar{r}  $ across the rows are: IMF $ \approx 0.40, $ SWS, $ \approx 0.61$, Kp $ \approx 0. 33$, Dst  $\approx 0.46 $, ap $ \approx 0.42$.
\end{figure*}

In examining each parameter, it is notable that SWS has the most significant impact on CR intensity reduction across the NMs, with an average correlation coefficient of $ r \approx $ 0.61.
This correlation coefficient value indicates that the magnitude of SWS results in a CR intensity reduction of about 60\% on the  Earth's surface, thus making high SWS a significant contributor to large SFds globally.
The geomagnetic indices have moderate correlation with the SFds, the values of their correlation coefficients are not strong enough to account for the SFds.  
They only showed that there is moderate geomagnetic activity on the days of the SFds.
This is expected as it is not likely to expect much disturbances in Earth's magnetosphere at a time when there is a large reduction in the intensity of CR reaching the Earth.  
The interesting thing is that in each of the NMs, the observables are similar as indicated by the range of the correlation coefficient.
The magnitude of the correlation coefficient being consistent across the NMs shows that the Earth's atmosphere is at similar conditions throughout the days of the SFds,

On the other hand, it is noteworthy that the average correlation coefficient regarding IMF across the stations being $ \approx $ 0.4, (accounting for a 40\% reduction in CR count reaching the Earth's surface) has a salient implication.  
This is because when the correlation coefficient of SFd-SWS and that of SFd-IMF is combined\footnote{i.e. summing up the effects of the two quantities}, the resultant value is $ r \approx $ 1.
The implication of this is that strong Solar Wind is primarily responsible for the large SFds as both SWS and IMF are attributes of the Solar Wind.
 Therefore, the intensity of the Solar Wind accounts for approximately 100\% of the SFd.

\section{Summary and Conclusion}

The interplay between the Fds, the interplanetary parameter, and geomagnetic indices across the NMs was investigated to ascertain if the interplay is simultaneous across the NMs.
In that, we are poised that if Fds are a global phenomenon, the effect of interplanetary parameters on the CR count in one station should be similar in all other stations if the Fds are simultaneous indeed.  
That we confirmed.
 
 From this current SFd catalogue it is seen that large SWS is responsible for the global reduction in CR intensity reaching the Earth's surface.
The IMF has no strong influence on CR reduction but plays a complementary role.  
The effects of the interplanetary parameters on the magnitude of the SFds are consistent across the NMs likewise the state of geomagnetic activity.
As such, we point out that testing the consistency of effects of interplanetary and geomagnetic on Fd across many NMs located at different points on Earth is another means of measuring strong or simultaneous Fds. We also note that the consistent results obtained for Fds and Dst and other solar-terrestrial agents are insightful as they can stimulate further research into the currently debated time delay between Fds and Dst in connection with their common solar-causing sources.

We note that one of the major flaws of the Fd list created here is the presence of CR anisotropy in the CR data. Conclusions drawn from the current work are tentative. Definitive inferences can only be reached when the analysis is repeated after removing the contribution from CR diurnal anisotropy. We will pursue this in future work.   
 
\section {Acknowledgments}
We obtained the data for this work at no cost from data. html hosted on http://cr0.izmiran.rssi.ru/ and https: //omniweb.gsfc. nasa.gov/html/ow, and we are grateful to their team.

Eya, I. O. expresses gratitude to the PanAfrican Planetary and Space Science Network (PAPSSN) Staff Mobility Scheme for enabling him to visit Copperbelt University (CBU). His visit to CBU provided him with ample opportunity to concentrate and complete this paper. The Astrophysics team at CBU also appreciates PAPSSN for providing this great opportunity that opened a door of collaboration between them and the Astronomy and Astrophysics research team at University of Nigeria Nsukka.
This paper makes use of the following websites: http://cr0.izmiran.rssi.ru/ and https://omniweb. gsfc.nasa.gov/html/owdata.html in sourcing our data. We appreciate the teams hosting these websites. 

The contribution from the reviewers of the article greatly changed the manuscript. We feel indebted to them.

\label{lastpage}

\end{document}